\documentclass[reprint,aps,prb,amsfonts,amssymb,amsmath,floatfix]{revtex4-2}
\pdfoutput=1
\usepackage[pdftex]{graphicx}
\usepackage[utf8]{inputenc}

\begin{document}

\title{Ultrafast Demagnetization Dynamics Due to Electron-Electron Scattering and Its Relation to Momentum Relaxation in Ferromagnets}

\author{Svenja Vollmar}
\author{Kai Leckron}
\author{Hans Christian Schneider}
\email{hcsch@physik.uni-kl.de}
\affiliation{Physics Department and Research Center OPTIMAS, TU Kaiserslautern, P. O. Box 3049, 67663 Kaiserslautern, Germany}

\date{\today}

\begin{abstract}
	We analyze theoretically the demagnetization dynamics in a ferromagnetic model system due to the interplay of spin-orbit coupling and electron-electron Coulomb scattering. We compute the $k$-resolved electronic reduced spin-density matrix including precessional dynamics around internal spin-orbit and exchange fields as well as the electron-electron Coulomb scattering for densities and spin coherences. Based on a comparison with numerical solutions of the full Boltzmann scattering integrals, we establish that the $k$-resolved reduced spin-density matrix dynamics are well described using a simpler generalized relaxation-time ansatz for the reduced spin-density matrix. This ansatz allows one to relate the complicated scattering dynamics underlying the demagnetization dynamics to a physically meaningful momentum relaxation time~$\tau$. Our approach reproduces the behaviors of the demagnetization time $\tau_{\text{m}} \propto 1/\tau$ and  $\tau_{\text{m}} \propto \tau$ for the limits of short and long $\tau$, respectively, and is also valid for the intermediate regime. The ansatz thus provides a tool to include the correct demagnetization behavior in approaches that treat other contributions to the magnetization dynamics such as transport or magnon/phonon dynamics.  
\end{abstract}

\maketitle

\section{Introduction}  

The relaxation of electrically or optically induced electronic spin polarizations in semiconductors and simple metals has been studied for more than 50 years and has important connections to the spin-dependent dynamics of electrons in ferromagnets. Spin relaxation dynamics in semiconductors have often been interpreted in terms of three different ``classical'' mechanisms: Elliott-Yafet(EY), Dyakonov-Perel(DP), and Bir-Aronov-Pikus, which were invented, respectively, for semiconductors with degenerate bands, for electronic bands with small spin splitting and small-band gap systems with electron-hole exchange interactions, see Refs.~\cite{zutic_spintronics_2004,fabian_semiconductor_2007} for a general overview.  The most widely applicable EY and DP mechanisms were based originally on a combination of spin-orbit coupling with electron-impurity and electron-phonon scattering.
In semiconductor spintronics, it was realized about 20 years ago that Coulomb scattering, which arises from the \emph{spin-independent} interaction that does not directly couple electrons to the lattice via phonons or impurities, can also contribute to spin relaxation. More precisely, it leads to spin dephasing in the presence of a $k$-dependent spin-orbit induced splitting between $\uparrow$ and $\downarrow$ states, which can be described in terms of a $k$-dependent internal effective magnetic field~\cite{wu_dyakonov-perel_2000,glazov_dyakonovperel_2003}. 

In ferromagnetic metals, a pronounced quenching of the magnetization, which is mainly related to d-band electrons, can be observed after excitation with an ultrashort optical pulse. While this is a more sizable effect than the relaxation of an induced spin polarization of a small density of excited electrons in semiconductor s- or p-like bands, the concept of Elliott-Yafet spin dynamics via electron-phonon scattering~\cite{yafet_g_1963} was introduced early on as a mechanism to explain the reduction of spin angular momentum observed in the demagnetization process of ferromagnets~\cite{koopmans_explaining_2010}.

The present paper is concerned with the characteristics of magnetization dynamics that are caused by a ``spin-relaxation like'' approach to magnetization dynamics. Compared to semiconductors, ferromagnetic metals possess a more complicated ground state with correlated d-electron bands at the Fermi level, more complicated elementary excitations (magnons) and arguably a different electron-phonon coupling (spin-lattice coupling). The mechanism of incoherent electronic dynamics~\cite{kraus_ultrafast_2009,mueller_feedback_2013} together with spin-orbit coupling considered in this paper thus competes with or complements other mechanisms, such as coherent electronic dynamics~\cite{dewhurst_angular_2021} coupling Fermi-level electrons to more tightly bound orbitals, direct angular momentum transfer to phonons~\cite{maldonado_tracking_2020,tauchert_polarized_2022} and magnon interactions~\cite{eich_band_2017,beens_modeling_2022}, to name only a few. 
The dominant scattering mechanisms contributing to the incoherent electron dynamics arise from the interaction with phonons and other electrons. Electron-phonon scattering is often regarded as important because it can lead to electronic spin flips via coupling to the lattice,~\cite{koopmans_explaining_2010} which fits into a picture of a three-temperature model as the spin-lattice coupling. Theoretical calculations indicate that the spin-dependent electron-phonon interaction, in which the phonons directly change electron spin, gives only a small contribution to electronic dynamics~\cite{essert_electron-phonon_2011,carva_ab_2011}. Instead, the main impact of electron-phonon scattering is related to providing an essentially spin-independent momentum scattering process that affects the spin in combination with electronic precessional spin dynamics~\cite{leckron_ultrafast_2017}. If the electron-phonon scattering contributes to magnetization dynamics mainly because it acts as a momentum scattering channel for electrons, then the electron-electron scattering provides an additional momentum scattering channel that should be even more important for highly excited electrons because it can act on an even shorter timescale of 10 femtoseconds. The demagnetization dynamics corresponding to the latter mechanism have so far been investigated at the level of Fermi's Golden Rule rates for transitions between spin-mixed states due to the Coulomb interaction~\cite{kraus_ultrafast_2009,mueller_feedback_2013}. This approach can explain a sizable contribution to demagnetization, in particular, if a dynamical Stoner exchange splitting is included~\cite{mueller_feedback_2013,leckron_ferromagnetic_2019}.  

In this paper we investigate the electron-electron scattering contribution to the spin-dependent dynamics in a ferromagnetic model system using a similar approach as we have employed for electron-phonon scattering~\cite{leckron_ultrafast_2017}. That is, we go beyond Fermi's Golden Rule rates for Coulomb scattering between electronic distributions in $k$-space and include the \emph{precessional dynamics} of coherences, i.e., the off-diagonal components of the spin-density matrix, around anisotropic effective spin-orbit fields.

Using a screening parameter to control the strength of the electron-electron Coulomb scattering, we find that the influence of this scattering mechanism, including its effects on the precessional dynamics, can be captured well using an extended relaxation time ansatz with a single effective momentum relaxation time~$\tau$ for any given interaction strength. The ansatz and the effective relaxation time provide an arguably more general description of relaxation processes than what can be obtained microscopically from our simple model band structure. In terms of this relaxation time we can consistently describe a whole range of different demagnetization behaviors from a proportionality to $\tau^{-1}$ to the proportionality to $\tau$, including the important intermediate regime, which has, to the best of our knowledge, not been mapped out in a ferromagnetic material yet. For semiconductors and non-magnetic metals, similar scalings of the spin relaxation rates/times have been found in their dependence of quasiparticle broadening~ \cite{boross_unified_2013,burkov_spin_2004} and doping concentrations~\cite{zhou_anomalous_2013}.



\section{Theoretical approach} 
Our theoretical approach to determine the demagnetization dynamics and the quantities involved in the electronic dynamics under the influence of internal spin-orbit fields and electron-electron scattering proceeds by first determining the single-particle energies and states of the Bloch electrons in our model band structure and then setting up and numerically solving the dynamical equations for the reduced electronic spin-density matrix including electron-electron Coulomb scattering. We then introduce here a relaxation time ansatz that can approximate the spin-conserving electron-electron Coulomb scattering well. The ansatz involves only a single relaxation time and introduces a time-dependent effective quasi-equilibrium spin-density matrix, to which the system evolves during demagnetization and remagnetization.
\subsection{Hamiltonian and Dynamical Equation}

The derivation of equations of motion (EOMs) for the electron-electron interaction is closely related to what we have presented in~Refs.~\onlinecite{baral_magnetic_2015} and \onlinecite{leckron_ultrafast_2017}, but the general approach has been well established for semiconductor spintronics earlier~\cite{wu_spin_2010}. Here, we only give a short overview of the model system, which uses a two-dimensional $k$-space to keep the single-particle band structure simple. The single-particle contribution to the system Hamiltonian is given by
\begin{equation}
	\hat{H}\left( \mathbf{k}\right)
	=\hat{H}_{\mathrm{kin}}\left( k\right) +\hat{H}_{\mathrm{SO}}\left( \mathbf{k}\right) +\hat{H}_{\mathrm{Stoner}}.
	\label{eq:hamiltonian-0}
\end{equation}
with the effective-mass contribution $\hat{H}_{\mathrm{kin}} \left( k\right)=\frac{\hbar^2k^2}{2m^*}$.
The spin-orbit contribution is assumed to be of the Rashba form and can be written in terms of the vector of Pauli matrices $\hat{\vec{\sigma}}$
\begin{equation}
	\hat{H}_{\mathrm{SO}}\left( \mathbf{k}\right) =\alpha \left( \hat{\vec{\sigma}} \times \mathbf{k} \right) \cdot \mathbf{e}_z=\alpha \left( \hat{\sigma}_x k_y -\hat{\sigma}_y k_x\right). 
\end{equation}
Its strength is controlled by the Bychkov-Rashba parameter~$\alpha$. The mean-field Hubbard contribution leads to a Stoner contribution $\hat{H}_{\mathrm{Stoner}} = Um$   
which depends on the magnetization $m$, see Eq.~\eqref{eq:spin_polarization_expectation_value}, and an effective on-site interaction energy $U$.

\begin{figure*}
	\includegraphics{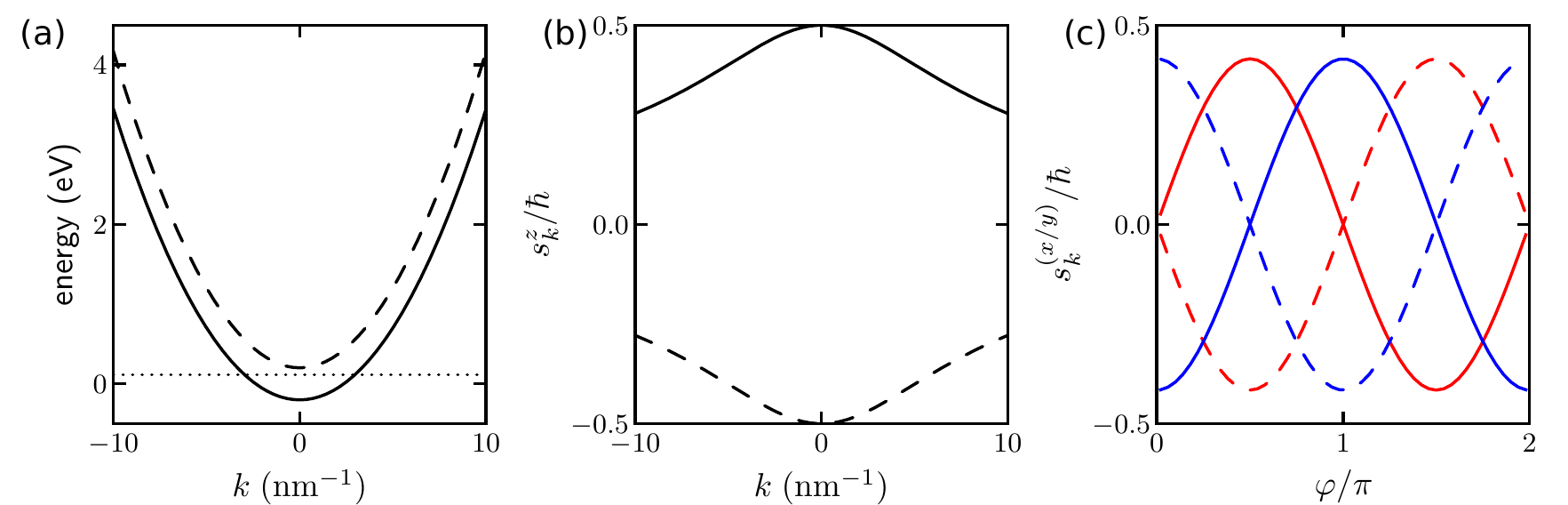}
	\caption{Self-consistently calculated $k$-dependent band structure (a) with the corresponding spin structure in $z$ direction (b). The $\varphi$-dependent spin structure in $x$ (red) and $y$ (blue) direction are shown for a fixed $k=10\;\mathrm{nm}^{-1}$ (c). The solid lines correspond to the energies and $k$-local quantization axes in the lower band, the dashed lines to those in the upper band. The parameters used throughout this paper are: Stoner interaction parameter $U = 400 \;\mathrm{meV}$, Rashba parameter $\alpha = 30\; \mathrm{meV nm^{-1}}$,  electron density $n_{\mathrm{e}} = 0.7\;\mathrm{nm}^{-1}$. At each $k$-point, the quantization axis of the upper and lower bands point in opposite directions.  \label{fig:energies-and-spins}}
\end{figure*}

Figure~\ref{fig:energies-and-spins} illustrates the important features of the model. The model band structure exhibits a $k$-dependent band splitting and $\mathbf{k}$-dependent Bloch spinors, which we denote by $\Uparrow$ and $\Downarrow$ to indicate that they are not pure spin states. The splitting $\Delta E_k \equiv \varepsilon_{k\Uparrow}-\varepsilon_{k\Downarrow}$ of the bands shown in Fig.~\ref{fig:energies-and-spins}(a) ranges from $\Delta E_{k=0}=400$\,meV to $\Delta E_{k=10\,\mathrm{nm}^{-1}}= 725$\,meV for different $k$-values. Fig.~\ref{fig:energies-and-spins}(b) and (c) depict the $k$-local spin expectation values $\langle \mathbf{k}\mu|\vec{\sigma}|\mathbf{k} \mu\rangle/2$ for the Bloch spinors $|\mathbf{k}\mu\rangle$: (b) the longitudinal component vs.~$k$ and (c) the components perpendicular to $z$ vs.~polar angle for a fixed $k$. Around $k=0$ states are essentially $\uparrow$ and $\downarrow$ spin states, but the mixing increases with increasing $k$, or equivalently, energy. For highly excited electrons at $k$-states as shown in Fig.~\ref{fig:energies-and-spins}(c), the Bloch states are considerably spin-mixed. 

The electronic quantum state is described by the reduced spin-density matrix $\rho_{\mathbf{k}}^{\mu\mu'} = \langle\hat{c}_{\mathbf{k}\mu}^{\dagger}\hat{c}_{\mathbf{k}\mu'}\rangle$, where $\hat{c}_{\mathbf{k}\mu}^{(\dagger)}$ is the annihilation (creation) operator of an electron with momentum $\mathbf{k}$ in band $\mu$. 
In the equation of motion (EOM) for the spin-density matrix, we include the electron-electron Coulomb interaction at the level of 2nd Born scattering integrals, which can be derived using Green function or reduced density-matrix techniques~\cite{binder_nonequilibrium_1995,haug_quantum_2008,kira_semiconductor_2012},
\begin{widetext}
\begin{equation}
	\begin{split}
		\frac{\partial}{\partial t} \rho_{\mathbf{k}}^{\mu \mu'}
		&=\frac{i}{\hbar}\left(\varepsilon_{\mathbf{k} \mu}
		-\varepsilon_{\mathbf{k}\mu'}\right)\rho_{\mathbf{k}}^{\mu \mu'}\\
		&+\frac{\pi}{\hbar}\sum_{\mathbf{l}\mathbf{q}} \sum_{\substack{\mu_1 \mu_2 \mu_3 \\ \mu_4 \mu_5 \mu_6 \mu_7}}
		\left(V_{\mathbf{k}\mathbf{l}\mathbf{q}}^{\mu \mu_1 \mu_2 \mu_3}\right)^*  
		\left( V_{\mathbf{k}\mathbf{l}\mathbf{q}}^{\mu_4 \mu_5 \mu_6 \mu_7} - V_{\mathbf{l} +\mathbf{q}\mathbf{l} \mathbf{k}-\mathbf{l}}^{\mu_4 \mu_5 \mu_6 \mu_7}\right)
		\delta \left(\Delta E^{\mu_4 \mu_5 \mu_6 \mu_7}_{\mathbf{k}\mathbf{l}\mathbf{q}}\right)\\
		& 
		\left[\rho_{\mathbf{k}+\mathbf{q}}^{\mu_3\mu_7 }\rho_{\mathbf{l}}^{\mu_2\mu_6} \left(\delta_{\mu_1\mu_5}{-} \rho_{\mathbf{l}+\mathbf{q}}^{\mu_5\mu_1}\right) 
		\left(\delta_{\mu'\mu_4}{-} \rho_{\mathbf{k}}^{\mu_4\mu'}\right)
		{-}\rho_{\mathbf{k}}^{\mu_4\mu'}\rho_{\mathbf{l}+\mathbf{q}}^{\mu_5\mu_1} \left(\delta_{\mu_2\mu_6}{-} \rho_{\mathbf{l}}^{\mu_2\mu_6}\right) 
		\left(\delta_{\mu_3\mu_7}{-} \rho_{\mathbf{k}+\mathbf{q}}^{\mu_3\mu_7}\right)\right]
		\\
		&+\frac{\pi}{\hbar}\sum_{\mathbf{l}\mathbf{q}} \sum_{\substack{\mu_1 \mu_2 \mu_3 \\ \mu_4 \mu_5 \mu_6 \mu_7}}
		V_{\mathbf{k}\mathbf{l}\mathbf{q}}^{\mu' \mu_1 \mu_2 \mu_3}
		\left( V_{\mathbf{k}\mathbf{l}\mathbf{q}}^{\mu_4 \mu_5 \mu_6 \mu_7} - V_{\mathbf{l} +\mathbf{q}\mathbf{l} \mathbf{k}-\mathbf{l}}^{\mu_4 \mu_5 \mu_6 \mu_7}\right)^* 
		\delta \left( \Delta E^{\mu_4 \mu_5 \mu_6 \mu_7}_{\mathbf{k}\mathbf{l}\mathbf{q}}\right)\\
		&
		\left[\rho_{\mathbf{k}+\mathbf{q}}^{\mu_7 \mu_3}\rho_{\mathbf{l}}^{\mu_6 \mu_2} \left(\delta_{\mu_5 \mu_1}{-} \rho_{\mathbf{l}+\mathbf{q}}^{\mu_1 \mu_5}\right) 
		\left(\delta_{\mu_4 \mu}{-} \rho_{\mathbf{k}}^{\mu \mu_4}\right)
		{-}\rho_{\mathbf{k}}^{\mu \mu_4}\rho_{\mathbf{l}+\mathbf{q}}^{\mu_1 \mu_5} \left(\delta_{\mu_6 \mu_2}{-} \rho_{\mathbf{l}}^{\mu_6 \mu_2}\right) 
		\left(\delta_{\mu_7 \mu_3}{-} \rho_{\mathbf{k}+\mathbf{q}}^{\mu_7 \mu_3}\right)\right].
	\end{split}
	\label{eq:EOM}
\end{equation}
\end{widetext}
The first row describes a coherent precession of the off-diagonal contributions of the spin-density matrix, i.e., the \emph{coherences} $\rho_{\mathbf{k}}^{\mu\mu'}, \mu\neq\mu'$, due to the splitting between the bands $\mu$ and $\mu'$ at $\mathbf{k}$. The remaining terms are electron-electron scattering contributions with the Coulomb-matrix elements $V_{\mathbf{k}\mathbf{l} \mathbf{q}}^{\mu_1	\mu_2 \mu_3 \mu_4}=V_q\left\langle\mathbf{k}\mu_{1}|\mathbf{k}+\mathbf{q}\mu_{4}\right\rangle\left\langle\mathbf{l}+\mathbf{q}\mu_{2}|\mathbf{l}\mu_{3}\right\rangle$ where $V_q$ denotes a \emph{screened} Coulomb potential depending on the momentum $\mathbf{q}$ transferred from the electron with initial momentum $\mathbf{k}$ to the electron with final momentum $\mathbf{l}$, i.e., $\mathbf{k}\rightarrow\mathbf{k}+\mathbf{q}$ and $\mathbf{l}+\mathbf{q}\rightarrow\mathbf{l}$. To obtain the Boltzmann-like scattering integrals in Eq.~\ref{eq:EOM} one has to employ a Markov approximation not only for real occupation-number distributions but also for complex coherences with a precessional contribution stemming from the first term. This precessional frequency is removed by transforming to a rotating frame, in which the Markov approximation can be made, and then transforming back.~\cite{wu_spin_2010,baral_magnetization_2014,Svenja_Diss} 

Since the $\mathbf{k}$-space for the single-particle states defined above is 2-dimensional, we use the screened Coulomb potential in two dimensions in the form $V_q=\frac{e^2}{2\mathcal{V}\varepsilon_0 \varepsilon_{\mathrm{b}} \left(q + \kappa \right)}$ ~\cite{haug_quantum_2009} with the elementary charge $e$, the normalization volume $\mathcal{V}$, the dielectric constant $\varepsilon_0$, the screening constant $\varepsilon_b$ and the screening parameter $\kappa$. This is not an essential restriction of our dynamical approach to 2-dimensional physics, it serves here only to simplify the numerical calculations, in particular the sums over momenta $\mathbf{l}$ and $\mathbf{k}$ on the RHS of Eq.~\eqref{eq:EOM}, which have to be calculated in every time step. Below we analyze the dependence of the dynamics in dependence of the inverse screening length, which is determined by the band structure, carrier density, and possibly by the dielectric environment. In our simplified band structure we essentially regard this as a model parameter and choose values on the order of $20\;\mathrm{nm}^{-1}$ and smaller. This value is consistent with a calculation of the screening parameter~$\kappa$ for electrons in parabolic bands~\cite{haug_quantum_2009,binder_nonequilibrium_1995} in the 2d-limit via $\kappa =m^* e^2/(2 \pi \hbar^2 \varepsilon_{\text{b}}\varepsilon_0)f(k=0)$ with relative background screening constant~$\varepsilon_{\text{b}}=1$.

One goal of this paper is to compare and contrast the electronic dynamics described by the full density-matrix with those obtained using \emph{occupation numbers}. In principle, an approach that uses only occupation numbers $n_{\mathbf{k}}^{\mu} := \rho_{\mathbf{k}}^{\mu\mu}$, i.e., the diagonal elements of the density matrix, is an approximation to the full density matrix. In this case, the Boltzmann scattering integrals are essentially rates as one would obtain from Fermi's Golden Rule, which connect non-pure spin states and thus lead to spin-flip transitions~\cite{steiauf_elliott-yafet_2009,koopmans_explaining_2010}. 
Because the Coulomb interaction is spin-independent, the Coulomb scattering \emph{alone} cannot cause a transition that changes the magnetization. For electron scattering dynamics between non-pure spin states, the restriction to occupation dynamics, which neglects the influence of the off-diagonal parts of the spin-density matrix, there is no conservation of ensemble spin and one obtains demagnetization due to Coulomb scattering~\cite{kraus_ultrafast_2009,mueller_feedback_2013}. In order to elucidate this,  
we also evaluate Eq.~\eqref{eq:EOM} at the level of Fermi's Golden Rule for occupations. This approach is often called ``Boltzmann scattering'', but this may cause confusion in our case because we also have Boltzmann-like scattering integrals for \emph{all} elements of the spin-density matrix in the complete EOM~\eqref{eq:EOM}. To differentiate between the full spin-density matrix calculation and the calculation that uses only the occupations, we refer to them as ``generalized Boltzmann scattering'' (or simply ``full'') and ``occupation-number approximation'', respectively.

The numerical solution of the EOM~\eqref{eq:EOM} requires a considerable accuracy to keep the numerical errors from accumulating over the demagnetization and remagnetization dynamics, which would spoil the important conservation laws. We thus use a Runge-Kutta-type integration method developed by Dormand and Prince~\cite{dormand_family_1980} with a dynamical time-step control to keep a high accuracy while also optimizing the CPU time.

\subsection{Relaxation-time approximation}\label{sec:relaxation-time}
In addition to the dynamics of the spin-density matrix with generalized Boltzmann scattering, which is non-local in $k$-space, we will use a relaxation-time ansatz that is designed specifically for spin-polarized systems with spin-orbit coupling. In Ref.~\cite{scholl_off-resonant_2019} we applied the ansatz in the context of optically driven dynamics mainly to simplify the numerical effort. Here, we stress that it allows one to replace the complexity of the scattering integrals by introducing a single physically meaningful relaxation time that characterizes the complex, $k$-dependent scattering dynamics. It thus provides a simple and intuitive, but also accurate description of this scattering process that should also have applications to calculations involving transport and/or in combination with other scattering mechanisms, such as electron-magnon scattering. 

The ansatz is based on suitably defined quasi-equilibrium density matrices of the general form
\begin{equation}
    \tilde \rho_{\mathrm{eq}} = f(T, \mu, \zeta_z)
    \label{eq:quasi_equilibrium_distribution}
\end{equation}
where $f$ is a Fermi-Dirac distribution depending on temperature~$T$, chemical potential~$\mu$ and spin accumulation~$\zeta$. The parameters~$T$,~$\mu$ and~$\zeta$ are determined such that the distribution reproduces a prescribed charge density, energy density and spin polarization. In this work, we only consider the spin polarization in $z$-direction due to the symmetries of our model system. The grand canonical Hamiltonian for non-interacting electrons corresponding to Eq.~\eqref{eq:quasi_equilibrium_distribution} is
\begin{equation}
    \hat{K} = \hat{H} + \mu \hat{N} - \zeta_{z} \hat \sigma_{z} \ ,
\label{eq:grand_canonical_hamiltonian}
\end{equation}
where $\hat{H}$ is the many-particle Hamiltonian corresponding to~\eqref{eq:hamiltonian-0} discussed above and $\hat{N}$ is the particle number operator.

The expectation values of the particle density, the spin polarization and the energy density used in the quasi-equilibrium distribution~\eqref{eq:quasi_equilibrium_distribution} will be obtained from those of the non-equilibrium density matrix $\rho$ as it arises during the dynamics and are calculated as follows. The electron density is given by
\begin{equation}
	n_{\mathrm{e}} = \frac{1}{\mathcal{V}}\sum_{\mu}\sum_{\mathbf{k}} \rho_{\mathbf k}^{\mu \mu}
    \label{eq:particle_density_expectation_value}
\end{equation}
the spin polarization/magnetization $m$ by
\begin{equation}
	m = \frac{1}{\mathcal{V}n_{\mathrm{e}}}\sum_{\mu \mu'}\sum_{\mathbf{k}} \left\langle \mathbf{k}\mu|\hat{s}_z|\mathbf{k}\mu'\right\rangle\rho_{\mathbf{k}}^{\mu \mu'}
	\label{eq:spin_polarization_expectation_value}
\end{equation}
and the energy density $\varepsilon$ by 
\begin{equation}
	\varepsilon = \frac{1}{\mathcal{V}}\sum_{\mu}\sum_{\mathbf{k}} \varepsilon_{\mathbf{k}\mu}\rho_{k}^{\mu \mu'}.
\label{eq:energy density_expectation_value}
\end{equation}
Here, $\mathcal{V}$ is the normalization volume. Note, in particular, that the spin-polarization dynamics also include the microscopic coherences $\rho_{\mathbf{k}}^{ \mu \mu'},\; \mu\neq\mu',$ and that, since we will only be discussing the relative change of the spin polarization further below, \emph{magnetization} and \emph{spin polarization} are interchangeable.  

Our relaxation-time approximation consists of replacing the scattering integrals by the following contribution to the equation of motion for the spin-density matrix
\begin{equation}
    \frac{\partial}{\partial t}\rho_{\mathbf{k}}^{\mu\mu'}\Big|_{\mathrm{rel}} = -\frac{\rho_{\mathbf{k}}^{\mu \mu'}-\tilde{\rho}_{\mathbf{ k}}^{\mu \mu'}}{\tau}.
    \label{eq:rta-dynamics}
\end{equation}
Here, $\tilde{\rho}_{\mathbf k}^{\mu \mu'}$ is the \emph{quasi-equilibrium} reduced spin-density matrix introduced in Eq.~\eqref{eq:quasi_equilibrium_distribution}, which is diagonal in the eigenbasis of the grand-canonical single-particle hamiltonian $\hat{K}$, see Eq.~\eqref{eq:grand_canonical_hamiltonian}. The dynamics of the spin-density matrix $\rho_{\mathbf{k}}^{\mu\mu'}$, however, is calculated in the eigenbasis of the regular single-particle Hamiltonian, so that one must transform the density matrix accordingly, i.e., from the eigenbasis of $\hat{K}$ to that of $\hat{H}$. Due to the transformation between the $\hat{K}$ and $\hat{H}$ bases,  $\tilde{\rho}_{\mathbf{k}}^{\mu \mu'}$, which is diagonal in the grand-canonical basis, has off-diagonal elements in the basis of $\hat{H}$ and therefore also describes the influence of scattering processes on the off-diagonal elements of the spin-density matrix, which are needed for the correct determination of the ensemble spin expectation value, i.e., the magnetization. With this approach, our Eq.~\eqref{eq:rta-dynamics} employs only a single relaxation time $\tau$ and mimics incoherent electron-electron scattering as it conserves the respective conservation laws. Note that the relaxation time is by construction independent of $k$ and energy and acts in a different way compared to relaxation times that are usually introduced as energy-dependent out-scattering rates and then averaged over suitably chosen quasi-equilibrium distribution functions~\cite{song_spin_2002,yu_spin_2005}.  Such an ansatz can also model spin-conserving electron-phonon scattering if one uses a different quasi-equilibrium distribution with a fixed temperature of the phonon bath $T_{\text{pn}}$. In this case, one only needs to determine $\mu$ and $\zeta$ in order to conserve density and spin polarization.

\subsection{Initial conditions and model excitation}\label{subsec:inital-excitation}

The magnetization dynamics discussed below start from a magnetic equilibrium state and since the Stoner contribution of the Hamiltonian~\eqref{eq:hamiltonian-0} depends on the spin-polarization $m$ of the system, this equilibrium state is determined self-consistently as follows: We start from an arbitrary value of $m$ in $z$-direction to set the preferential direction and iteratively (i) calculate the new band structure according to $m$, (ii) populate this band structure with electronic equilibrium (Fermi) distributions
\begin{equation}
	\rho_{k}^{\mu \mu'}=\frac{1}{e^{\beta(\varepsilon_{k \mu}-\mu )}+1} \delta_{\mu \mu'}.
	\label{eq:rho-FD}
\end{equation}
by adjusting the chemical potential $\mu_{\mathrm{C}}$ such that our desired electronic density $n_\mathrm{e}$ for a given equilibrium temperature $T_\mathrm{eq}$ is reproduced and (iii) calculate the new spin polarization $m_{\mathrm{new}}$ in this band structure and repeat steps (i)--(iii) until the spin polarization difference $\Delta m$ between two consecutive iterations is small enough (we chose $\Delta m <10^{-9}$).

In order to achieve a comparison of the different magnetization dynamics we employ a simple model excitation by an ultrashort optical pulse. As before~\cite{leckron_ultrafast_2017,vollmar_generalized_2017} we assume that the electronic energy is instantaneously raised to an excitation temperature~$T_{\mathrm{ex}} \gg T_{\mathrm{eq}} = 100\;\mathrm{K}$ and the excited electrons are distributed according to Eq.~\eqref{eq:rho-FD} with the excitation temperature in the self-consistently determined band structure. 

The instantaneous heating leads to different chemical potentials for the $\Uparrow$ and $\Downarrow$ bands as well as to a small change of the spin polarization $m$ due to the $\mathbf{k}$-dependent spin-mixing of the states. This approach does not change the electronic density in each band and is numerically simple and controllable by a single parameter $T_{\mathrm{ex}}$; it is designed to capture qualitatively the effect of an ultrashort pulse that deposits energy in the electronic system, see, e.g., Ref.~\cite{essert_electron-phonon_2011} for a more detailed description of this process. With this model of the excitation process we neglect optically driven interband coherences that may be excited by the excitation pulse, as the purpose of this paper is to analyze the possible dynamics in the incoherent regime.

We also choose a high excitation temperature $T_{\mathrm{ex}}$ of 4000\;K in order to clearly exhibit the dynamical features. The temperature~$T_{\mathrm{ex}}$ is purely a measure of the excitation of the electronic system and is well above the Curie temperature, which is $T_{\mathrm{C}} \approx 1030 \; \mathrm{K}$ for the model parameters listed in Fig.~\ref{fig:energies-and-spins}.

\section{Results\label{sec:results}}

We start with the generalized Boltzmann scattering. The main quantity of interest to us is the time-dependence of the magnetization $m$, i.e., the spin polarization of the electrons in the split bands. As it is our goal to compare the dynamics for different ratios of typical scattering times to typical precession times, we choose here to vary \emph{only} the screening parameter $\kappa$, which changes the matrix element of the Coulomb interaction and thus the $k$-dependent scattering rates. Using this one adjustable parameter to control the Coulomb scattering rates, we intend to illustrate the range of possible behaviors of the magnetization dynamics and how well these can be captured with our extended relaxation time ansatz. We therefore make no effort here to connect $\kappa$ to a variable that can be tuned in experiment.

\begin{figure}
	\includegraphics{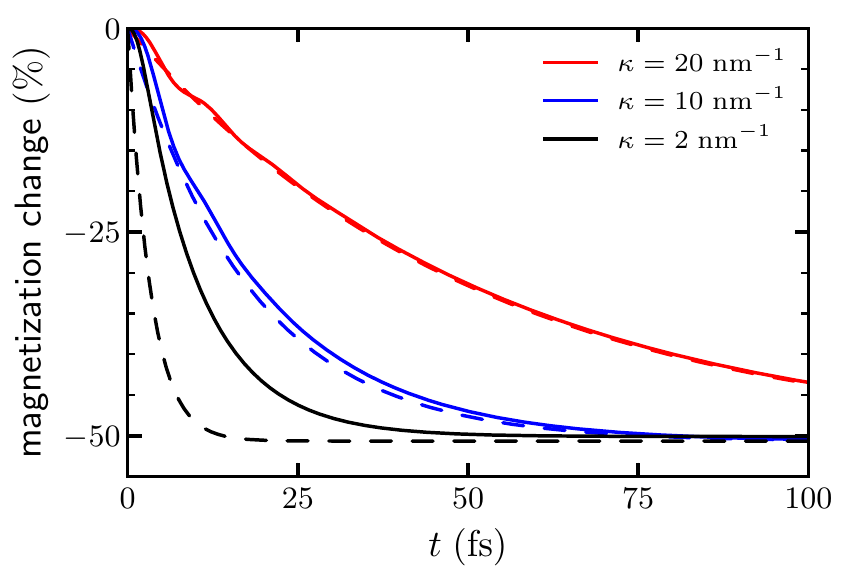}
	\caption{Relative magnetization change vs. time for different screening parameters $\kappa$ for the full calculation (solid lines) and the occupation-number  approximation (dashed lines). }
	\label{fig:kappa}
\end{figure}

\begin{figure}
	\includegraphics{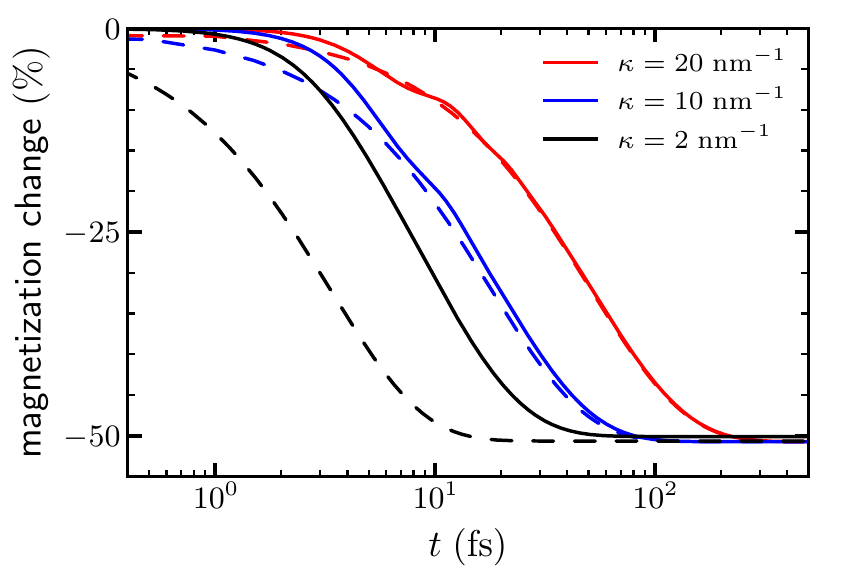}
	\caption{Same data as in Fig.~\ref{fig:kappa} displayed with a logarithmic time axis to emphasize the short-time dynamics. Regardless of $\kappa$, there is only a marginal magnetization change for the full calculation (solid lines) on short times $<4$\,fs since the build-up of the coherences is limited by the precession time of $\approx 10$\;fs. The demagnetization process in the occupation-number approximation starts at arbitrarily early times for stronger scattering, i.e., smaller $\kappa$.}
	\label{fig:kappa-log}
\end{figure}

Figure~\ref{fig:kappa} shows the demagnetization dynamics obtained for the excitation conditions discussed in Sec.~\ref{subsec:inital-excitation}. Shown are the relative magnetization changes resulting from the full calculation (solid lines) and the calculation using the occupation-number approximation (dashed lines) for small, intermediate and large screening parameters $\kappa =2$\;nm$^{-1}$, $\kappa =10$\;nm$^{-1}$ and $\kappa=20$\;nm$^{-1}$. We do not include electron-phonon coupling, or any other coupling to an energy bath that would absorb the energy transferred to the electronic system by the optical excitation over time. The system therefore stays in the demagnetized state even for times $t\gtrsim 0.1\;$ps. For the occupation-number approximation a smaller screening parameter $\kappa$ (faster scattering) always leads to faster spin/magnetization dynamics. In the full calculation, this behavior is also visible but less pronounced and the magnetization dynamics only coincide for larger $\kappa$ (slower scattering) with those of the occupation-number approximation. For values of $\kappa \lesssim 5\; \mathrm{nm}^{-1}$ the two calculations deviate at shorter times and the demagnetization dynamics of the full calculation are considerably slower. This is shown more prominently in Fig.~\ref{fig:kappa-log}, which plots the same curves vs.~a logarithmic time scale to display the behavior at very short time scales. Around and below the precession time $T_{\mathrm{p}} = h/\Delta E_k \approx 10$ fs  one observes that $T_{\mathrm{p}}$ effectively sets a lower limit for any magnetization change by this mechanism. The reason for this discrepancy is that the occupation-number dynamics neither includes precessional spin dynamics nor its dephasing due to the Coulomb interaction. The Golden-Rule like scattering in the occupation-number dynamics simply gets faster for larger interaction matrix elements.

In an earlier paper~\cite{leckron_ultrafast_2017} we have studied the influence on spin dephasing of electron-longitudinal-phonon scattering, for which the typical momentum scattering rates are longer than the precession times of electronic spins around typical exchange fields (determined by the exchange splitting between the bands). For this mismatch of scattering and precession times, we found that the occupation-number approximation agreed well with the full calculation, and that it is thus justified to use an Elliott-Yafet like description, i.e., an incoherent scattering process that leads to a spin change, or, as it is often called, a spin flip, and exhibits a linear time-dependence of demagnetization times on typical electron-phonon scattering times. 
For the electron-electron scattering considered in this paper, the occupation-number approximation reproduces the result of the full calculation \emph{only} for stronger screening effects. 
The discrepancy between the occupation-number approximation and the full dynamics should be even more pronounced in systems with a smaller splitting, where the precession frequencies are smaller.

\begin{figure}
	\includegraphics{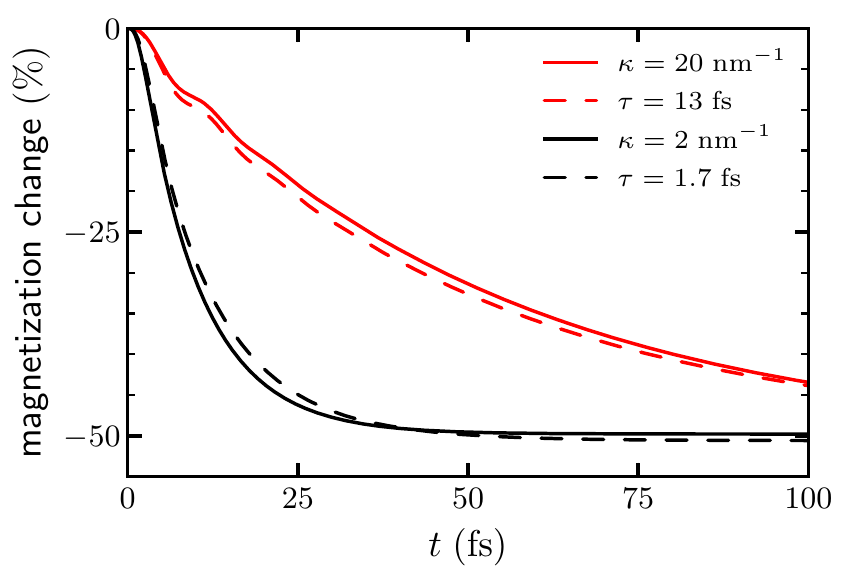}
	\caption{Relative magnetization change vs. time for the full calculation with two different screening parameters (solid lines) and two relaxation time calculations (dashed lines). The remaining parameters are as in Fig.~\ref{fig:kappa}.}
	\label{fig:kappaVSRelaxation}
\end{figure}

In Fig.~\ref{fig:kappaVSRelaxation} we turn to a comparison of the full calculation with the relaxation-time ansatz. We start with the demagnetization characteristics as obtained for two different screening parameters and suitably chosen relaxation times $\tau$ for each screening parameter. We obtain a good agreement of the demagnetization dynamics between the $\kappa = 20\;\mathrm{nm}^{-1}$ and $\tau = 13\;\mathrm{fs}$ curves as well as between the $\kappa = 2\;\mathrm{nm}^{-1}$ and $\tau = 1.7\;\mathrm{fs}$ curves. This is already a satisfying result, as a look back to Fig.~\ref{fig:kappa} shows that such a good agreement for different $\kappa$ values cannot be achieved using the occupation-number approximation.  However, the magnetization change is a  $k$-integrated quantity, and we would also like to compare the $k$-resolved dynamics. 

\begin{figure}
	\includegraphics{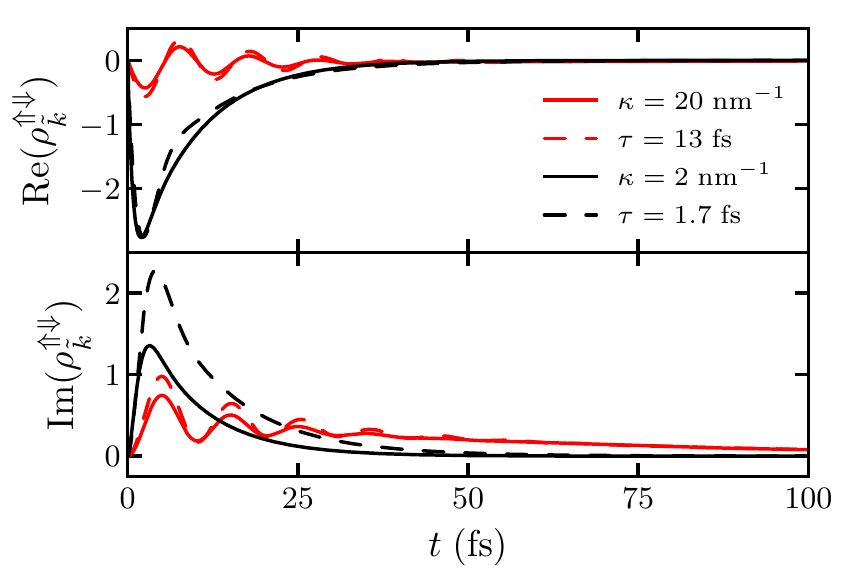}
	\caption{Real (top) and imaginary part (bottom) of the coherence at a $k$-point $\tilde{k} = 3.3\;\mathrm{nm}^{-1}$ near the Fermi edge vs time for the full calculation with two different screening parameters (solid lines) and two relaxation time calculations (dashed lines). The remaining parameters are as in Fig.~\ref{fig:kappa}.}
	\label{fig:kappaVSRelaxation-kohaerenz}
\end{figure}

Figure~\ref{fig:kappaVSRelaxation-kohaerenz} compares the results of the full calculation and the relaxation-time approximation for 
the dynamics of the coherence $\rho_k^{\Uparrow\Downarrow}$ at $\tilde{k}=3.3\;\text{nm}^{-1}$ for different values of $\kappa$. This particular $\tilde{k}$ is located near the Fermi edge of the lower band so that its dynamics play an important role during the whole demagnetization process. The different $k$-dynamics for the two cases that arise in the full microscopic calculation are reproduced well by the calculation with the extended relaxation time ansatz. For the full calculation and the relaxation-time ansatz precessional dynamics of the coherence are clearly visible at early times $t \lesssim 30\;\mathrm{fs}$ for $\kappa = 20\;\mathrm{nm}^{-1}$ and the fit with $\tau = 13\;\mathrm{fs}$. There are two contributions we want to discuss here. First, the precessional dynamics is driven by transitions $\mathbf{k}\to \mathbf{k}+\mathbf{q}$ which conserve the vector spin and thus lead to a mismatch of the spin with the local quantization axis at $\mathbf{k}+\mathbf{q}$. Second, the scattering also dephases the precession of the spin coherences toward a finite value at around $t=30\;\mathrm{fs}$, at which time the demagnetization is not even half completed (see Fig.~\ref{fig:kappa}). After that time, there is only a slow relaxation of the spin coherence. This result is qualitatively similar to that obtained for spin-conserving electron-phonon scattering~\cite{leckron_ultrafast_2017}. 

For $\kappa = 2\;\mathrm{nm}^{-1}$ in the full calculation, the respective $\tau = 1.7\;\mathrm{fs}$ that fits best in the relaxation time approach becomes shorter by a factor of 8, indicating a much faster scattering. In this case one cannot discern a precessional motion in the coherence, but rather strongly damped dynamics with one intermittent maximum. The whole spin-density dynamics here occurs on essentially the same time scale as the corresponding demagnetization curve in Fig.~\ref{fig:kappaVSRelaxation}. In both cases the relaxation-time ansatz comes close to the full calculation even though it replaces the complicated electron-electron scattering dynamics in $k$-space by a $k$-local expression with a constant, i.e., $k$-independent, relaxation time. That the quantitative agreement of the $k$-resolved and $k$-integrated quantities is so good is likely also due to our model excitation which creates excited electron distributions that are close to hot Fermi-Dirac distributions. For excitations that exhibit stronger non-equilibrium characteristics, the quantitative agreement may not be quite as good, but the relaxation time approximation should still capture the most important features of the spin-dependent electronic scattering dynamics.

\begin{figure}
	\includegraphics{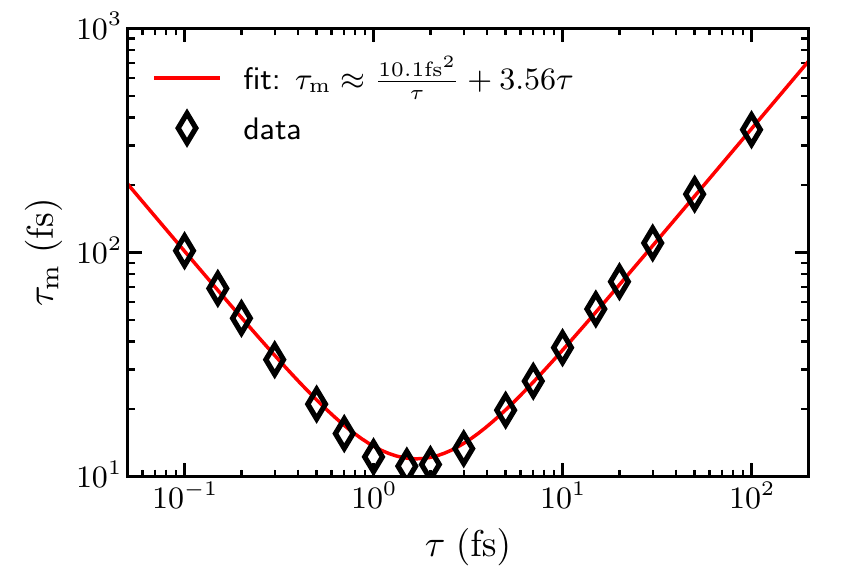}
	\caption{Demagnetization time vs effective scattering time $\tau$ with a fit. The remaining parameters are as in Fig.~\ref{fig:kappa}.}
	\label{fig:tauvariation}
\end{figure}

Up to now we have demonstrated that the relaxation-time ansatz reproduces the magnetization dynamics due to precessional dynamics and scattering quite well. Because the complicated scattering dynamics are parametrized by the relaxation time $\tau$ we can use it to study the influence of scattering on the electronic spin-dependent dynamics via this single parameter. For a range of relaxation times~$\tau$, we calculate the demagnetization curves as in Fig.~\ref{fig:kappaVSRelaxation} and subsequently extract the \emph{demagnetization time}~$\tau_{\mathrm{m}}$ by fitting the demagnetization dynamics by an exponential function $m(t) = b-a\exp(t/\tau_{\mathrm{m}})$. Fig.~\ref{fig:tauvariation} plots the demagnetization times obtained from this fitting procedure vs.\ the relaxation time~$\tau$ as black diamonds. For long relaxation times~$\tau$, the demagnetization time increases with~$\tau$, but for short relaxation times, they increase with \emph{decreasing} $\tau$. The scaling for small and large $\tau$ is in agreement with that found in Ref.~\onlinecite{zhou_anomalous_2013} for electronic dynamics in quantum wells. Importantly, there exists a minimum of $\tau_{\mathrm{m}}(\tau)$ for intermediate~$\tau$. Calculations using the occupation-number approximation (not shown) miss the behavior at small $\tau$ and yield only a steady decrease of $\tau_{\mathrm{m}}$ with decreasing relaxation times $\tau$. This behavior is already evident in Fig.~\ref{fig:kappa} where the occupation-number approximation leads to ever faster demagnetization.

The quantitative dependence of $\tau_{\mathrm{m}}$ on $\tau$ is fit to a function of the form 
\begin{equation}
    \tau_{\mathrm{m}} = \frac{A}{\tau} + B\tau\ ,
    \label{eq:tau-m-fit}
\end{equation}
which is  shown as red solid line in Fig.~\ref{fig:tauvariation}. We see a remarkably good agreement of the fit and the extracted demagnetization times from the relaxation time calculations. The parameter $A=10.1\;\mathrm{fs}^2$ can be put into correspondence with the precession frequency of the spin-splitting  $\hbar \Omega = \Delta E_k$ via $A\simeq \Omega^{-2}$. In the short~$\tau$ limit we thus obtain a connection $\tau_{\mathrm{m}} \approx \Omega^{-2}/\tau$. 
The form of Eq.~\eqref{eq:tau-m-fit} has already been suggested by  Refs.~\onlinecite{burkov_spin_2004,boross_unified_2013}. It can be obtained, for instance, from a thermal Green function approach to spin relaxation in semiconductors and metals. The relation of the spin relaxation time to a characteristic time $\tau_0=\gamma^{-1}$ in Refs.~\onlinecite{burkov_spin_2004,boross_unified_2013} is very similar to ours, but there the momentum scattering rates are related to lifetimes at the Fermi energy. Our results suggest that the scaling of the demagnetization time with $\tau$ is rather robust and also valid for excited systems with electronic populations far away from the Fermi energy if $\tau$ is interpreted as in Eqs.~\eqref{eq:quasi_equilibrium_distribution} and~\eqref{eq:rta-dynamics}. 

While the behavior for short $\tau$ is reminiscent of typical spin dephasing mechanisms of spintronics, the same microscopic interplay of precessional spin dynamics around internal fields with a spin-independent scattering mechanism is behind the demagnetization dynamics \emph{for the whole range} of $\tau$ shown in Fig.~\ref{fig:tauvariation}. In particular for larger~$\tau$ we obtain the inverse relation $\tau_{\mathrm{m}} \propto \tau$. Such a relation is usually associated with spin relaxation as it occurs via spin-flip transitions due to an explicitly spin-dependent interaction. 
Fig.~\ref{fig:tauvariation} shows clearly that both behaviors occur as limiting cases for small and large $\tau$, respectively, for electrons in a ferromagnetic band structure with spin-orbit coupling. For intermediate $\tau$ of about 1 to 10\;fs, a minimum  of demagnetization times occurs which is also very well described by the fit curve. The range around the minimum is likely close to the realistic range for metallic systems. We believe that the result contained in Fig.~\ref{fig:tauvariation} gives an accurate and intuitive picture of electron-electron scattering dynamics in highly excited ferromagnets by identifying $\tau$ as a physically meaningful parameter. One can thus use it as a fit parameter also for systems that are not described by the model band structure used in this paper. This makes it possible to extract $\tau$ from measured data $\tau_{\mathrm{m}}$ data via a fit, or obtain $\tau_{\mathrm{m}}(\tau)$ from numerical calculations solving the full Boltzmann scattering problem. Fig.~\ref{fig:tauvariation} is particularly important for fits to experimental $\tau_{\mathrm{m}}$ data. If one does not include the nonlinear regime at small~$\tau$ and assumes a linear relation between $\tau_{\mathrm{m}}$ and $\tau$ one would greatly overestimate the actual momentum relaxation time and miss the contribution from the precessional dynamics completely.

\section{Conclusion}

In this paper we discussed the spin-dependent incoherent carrier dynamics due to electron-electron scattering in a ferromagnetic model system with spin-orbit coupling, which provides an extension of our earlier study of electron-phonon scattering in this system. We described a dynamical calculation using the spin-density matrix, spin-orbit coupling and electron-electron scattering, which is non-local in $k$-space, and compared the numerical results with a calculation using only occupation numbers and with a generalized relaxation-time ansatz. We found that the calculation using only occupation numbers failed to capture the demagnetization behavior for weak screening, i.e., strong scattering because the precessional dynamics around spin-orbit fields is neglected. The comparison with the generalized relaxation-time ansatz showed a very good agreement both for weak and strong Coulomb scattering, i.e., in situations where precessional dynamics of the off-diagonal part of the reduced spin-density matrix are clearly visible and also in cases where they are suppressed by scattering/dephasing. This suggests that the relaxation-time ansatz can capture essential properties of the incoherent spin-dependent dynamics using a $k$-local expression with a \emph{single momentum relaxation time}~$\tau$. Such a simpler form  should be useful in numerical calculations for more complicated problems, in which scattering/dephasing due to the Coulomb interaction plays a role, such as transport and/or electronic dynamics due to coupling to magnons. In terms of the momentum relaxation time we were able to fit the calculated demagnetization times using a sum of terms proportional to $\tau$ and $\tau^{-1}$. The  $\tau_{\text{m}}$ vs.~$\tau$ relationship is much simpler than the microscopic electron dynamics and such a fit should be possible regardless of the details of the underlying band structure.

\begin{acknowledgments} 
This work was supported  by the Deutsche Forschungsgemeinschaft (DFG, German Research Foundation) - TRR 173/2 - 268565370 Spin+X (Project B03).
\end{acknowledgments}

%

\end{document}